\documentclass[10pt, conference,letterpaper]{IEEEtran}
\IEEEoverridecommandlockouts
\usepackage{authblk}
\usepackage{cite}
\usepackage{amsmath,amssymb,amsfonts,dsfont}
\usepackage{algorithm,algorithmic}
\usepackage{graphicx}
\usepackage{textcomp}
\usepackage{xcolor}
\usepackage{bm}
\usepackage{booktabs}
\usepackage{multirow}
\usepackage{lipsum}
\usepackage{url}
\usepackage{amsthm}
\usepackage{enumerate}
\usepackage[right = 0.66in, left = 0.66in,top=0.74in,bottom=1.05in]{geometry}
\ifCLASSOPTIONcompsoc
\else
\usepackage[caption=false, font=footnotesize]{subfig}

\newtheorem{lemma}{Lemma}
\newtheorem{remark}{Remark}

\newcommand{\tr}{\text{Tr}}


\begin{document}

\title{An Efficient Alternating Minimization Algorithm for Computing Quantum Rate-Distortion Function}
\author[1]{Lingyi Chen}
\author[1]{Deheng Yuan}
\author[2]{Wenyi Zhang}
\author[1]{Hao Wu}
\author[3$\dag$]{Huihui Wu}
\affil[1]{Department of Mathematical Sciences, Tsinghua University, Beijing 100084, P.R. China.}
\affil[2]{Department of Electronic Engineering and Information Science, 
\authorcr University of Science and Technology of China, Hefei, Anhui 230027, P.R. China. 
}
\affil[3]{Zhejiang Key Laboratory of Industrial Intelligence and Digital Twin,
\authorcr Eastern Institute of Technology, Ningbo, Zhejiang 315200, P.R. China.
\thanks{{$^{\dag}$}Corresponding author. Email: huihui.wu@ieee.org  (H. Wu).}
\thanks{This work was supported by the National Natural Science Foundation of China (Grant Nos. 12271289 and 62231022).}}


\maketitle

\begin{abstract}
We consider the computation of the  entanglement-assisted quantum rate-distortion function, which plays a central role in quantum information theory. 
We propose an efficient alternating minimization algorithm based on the Lagrangian analysis. 
Instead of fixing the multiplier corresponding to the distortion constraint, we update the multiplier in each iteration.
Hence the algorithm solves the original problem itself, rather than the Lagrangian relaxation of it.
Moreover, all the other variables are iterated in closed form without solving multi-dimensional nonlinear equations or multivariate optimization problems.
Numerical experiments show the accuracy of our proposed algorithm and its improved efficiency over existing methods.
\end{abstract}


\section{Introduction} \label{sec_introduction}
The entanglement-assisted quantum rate-distortion function~\cite{datta2013quantum,Wilde2013quantum}, or quantum rate-distortion function, is an important theoretical limit in quantum rate-distortion theory~\cite{Barnum2000Quantum,Devetak2002quantum,datta2013quantum,Wilde2013quantum}.
It addresses the trade-off between the compression of quantum data and the preservation of its fidelity, with the assistance of pre-shared entangled quantum states.
It also provides a lower bound for the unassisted quantum rate-distortion function, for which a single letter characterization is unknown except in some special cases~\cite{Devetak2002quantum,Datta2013quantumtoclassical,Chen2008Entanglement,Wilde2013quantum}.
Due to its simple form and clear physical meaning, the quantum rate-distortion function is relatively easy to analyze and offers insight into quantum lossy compression problems. 
It can be seen as an extension of the classical rate-distortion function, which is the compression limit of a classical source~\cite{shannon1959coding,berger1971} and plays a fundamental role in lossy data compression.

The significance of the quantum rate-distortion function aroused a lot interest in its calculation.
Some special cases like the isotropic qubit source admit analytical solutions~\cite{Wilde2013quantum}, however, for most examples closed form solutions have not been obtained. 
Thus, it is necessary to develop efficient numerical computation methods.
The quantum rate-distortion function involves a {\color{black}semi-definite constraint} in which the variables are typically required to be positive semi-definite matrices.
The complicated and highly non-linear semi-definite constraint brings much difficulty to the computation. 
{As a special case of }the quantum rate-distortion function, the classical rate-distortion function is efficiently solved by the well-known Blahut-Arimoto (BA) algorithm~\cite{blahut1972computation,arimoto1972algorithm} and more efficient algorithms have been developed recently~\cite{wu2022communication,chen2023constrained,hayashi2023bregman}.
However, there are only linear constraints and the difficult semi-definite constraint trivially degenerates into non-negativity constraints in the classical rate-distortion function, rendering its computation much easier. 
Standard methods for the classical rate-distortion function are not applicable to the computation of the quantum rate-distortion function, and new numerical methods for the latter are needed. 
 
{Although the quantum rate-distortion problem can be formulated as a {\color{black}non-linear convex program with a semi-definite constraint} and solved by general-purpose solvers~\cite{Chandrasekaran2017Relative,Fawzi2019Semidefinite,Coey2023Performance}, more efficient computation methods should exploit its specific structure.
Prior work~\cite{hayashi2023bregman}} derived EM algorithms for both the classical and quantum rate-distortion functions, based on a Bregman divergence framework in the general problem setting of information geometry.
In~\cite{he2024efficient} the mirror descent algorithm was extended to efficiently compute the quantum rate-distortion function, with the help of techniques such as symmetry reduction, duality, and inexact computation of iterations.
However, in algorithms in~\cite{hayashi2023bregman,he2024efficient} one of the alternating sub-problems does not have a closed form solution, hence solving such sub-problem requires an inner iteration for a multi-dimensional nonlinear equation system or a multivariate optimization problem.
The multiple layers of iterations in these algorithms incur a high cost and are not scalable with the problem size.
Moreover, fixing the Lagrangian multiplier of the distortion constraint and solving the resulting
Lagrangian relaxation of the original problem~\cite{he2024efficient} can lead to excessive computation cost as discussed in~\cite{wu2022communication,chen2023constrained}.

In this work, an efficient alternating minimization algorithm based on the Lagrangian analysis for computing the quantum rate-distortion function is proposed. Instead of fixing the multiplier, we update the multiplier in each iteration, thus, solving the original problem. By updating the multiplier, the computational efficiency can be {significantly improved, as demonstrated in classical cases \cite{wu2022communication,chen2023constrained,ye2022optimal}.} Moreover, all the other variables are iterated in closed form without solving multi-dimensional nonlinear equations or multivariate optimization problems, guaranteed by a lemma to solve a program involving the entropy function in closed form.
{Numerical experiments demonstrate that solutions generated by our algorithm have  high accuracy, and also reveal that our algorithm achieves better computational efficiency relative to existing methods. }
%

\section{Problem Formulation}

Denote the $n \times n$ identity matrix by $I_n$.
Denote the set of $n \times n$ Hermitian matrices by $\mathbb{H}^n$ and its subset of $n \times n$ positive semi-definite matrices by $\mathbb{H}^n_+$. Let $\mathcal{D}^n \triangleq \{\rho \in \mathbb{H}^n_+: \tr \rho = 1\}$, which is the set of $n \times n$ density matrices (i.e. positive semi-definite matrices with unit trace).
Let $\exp(\cdot )$ be the matrix exponential function and $\log(\cdot)$ be the matrix logarithmic function.
For $A \in \mathbb{C}^{n \times n}$ and $B \in \mathbb{C}^{m \times m}$, let $A \otimes B \in \mathbb{C}^{(mn) \times (mn)}$ be the Kronecker product of $A$ and $B$.
For each $C \in  \mathbb{C}^{(mn) \times (mn)} (= \mathbb{C}^{n \times n} \otimes \mathbb{C}^{m \times m})$, let $\tr_1(C) \in \mathbb{C}^{m \times m}$ and $\tr_2(C) \in \mathbb{C}^{n \times n}$ be the partial traces over the first system $\mathbb{C}^{n \times n}$ and over the second system $\mathbb{C}^{m \times m}$, respectively.
Let $\tr(\cdot)$ be the classical trace function.

Then the quantum rate-distortion function can be written (cf. Lemma 21 in~\cite{hayashi2023bregman}) as follows.
\begin{subequations}\label{QRD}
\begin{align}
     &R(D) \triangleq \min_{\substack{\rho_{RB} \in \mathbb{H}^{mn}_+,\\ \sigma_B \in \mathcal{D}^{m}} }S(\rho_{RB}\|\rho_R\otimes \sigma_B),  \label{QRDa}
     \\
     &\text{subject to } \tr(\Delta \rho_{RB})\leq D,\ \tr_2(\rho_{RB})=\rho_R, \label{QRDb}
\end{align}
\end{subequations}
given a density matrix $\rho_R \in \mathcal{D}^{n}$, an $(mn) \times (mn)$ positive semi-definite distortion matrix $\Delta \in \mathbb{H}^{mn}_+$ and a distortion criterion $D\geq 0$. Here, $S(\cdot\|\cdot)$ is the quantum relative entropy, \textit{i.e.},  
$$S(\rho\|\sigma)=\tr(\rho\log \rho)-\tr(\rho\log\sigma).$$
Moreover, let $\Delta_B = \tr_1(\Delta(\rho_R \otimes I_m))$ that is positive semi-definite, and $\lambda_{\min}(\Delta_B) \geq 0$ be the minimum eigenvalue of~$\Delta_B$.
By Lemma 21 in~\cite{hayashi2023bregman}, if $\lambda_{\min}(\Delta_B) \leq D$, then $R(D) = 0$; 
otherwise, in the problem~\eqref{QRD} the inequality constraint can be replaced by an equality, i.e. $\tr(\Delta \rho_{RB})= D$ in~\eqref{QRDb}.
Hence it is sufficient to focus on the case $\lambda_{\min}(\Delta_B) > D$ and address the minimization in~\eqref{QRD} with $\tr(\Delta \rho_{RB})= D$.

The problem~\eqref{QRD} is a \textcolor{black}{non-linear convex program} in which one of the variables $\rho_{RB}$ is under complicated constraints. 
Specifically, it is required to be a positive semi-definite matrix and to satisfy a series of linear constraints in~\eqref{QRDb} simultaneously.
{Maintaining the satisfaction of these complicated constraints directly can require solving multi-dimensional nonlinear equation systems or multivariate optimization problems repeatedly~\cite{hayashi2023bregman,he2024efficient}, thereby incurring substantial computational costs.
To overcome the difficulty posed by these constraints,} we introduce dual variables to eliminate the constraints in~\eqref{QRDb}, and update primal and dual variables in an alternating way in the next section.

\section{Alternating Minimization Algorithm}\label{algorithm}
To solve the quantum rate-distortion problem \eqref{QRD},  we introduce dual variables $\beta\in \mathbb{R}^+$ and $\Lambda_R\in \mathbb{R}^{n\times n} \cap \mathbb{H}^n$, and then the Lagrangian function of \eqref{QRD} is written as follows:
\begin{multline}
\label{Lagrangian}
\!\!\!\!\!\mathcal{L}(\rho_{RB},\sigma_B;\beta,\Lambda_R)\!=\!\tr(\rho_{RB}\log \rho_{RB})
    -\tr(\tr_1(\rho_{RB})\log \sigma_B)\\
    +\beta(\tr(\rho_{RB}\Delta)-D)-\tr(\Lambda_R(\tr_2(\rho_{RB})-\rho_R)).
\end{multline}
Here, $\rho_{RB}\in \mathbb{H}^{mn}$, $\sigma_B\in D^m$, \textcolor{black}{and we ignore the term $-\tr(\tr_2(\rho_{RB})\log \rho_R)$ since it is a constant under~\eqref{QRDb}.} 
Based on that, we minimize the primal variables to obtain the dual problems and then update dual variables in an alternating way, as described below:
\begin{itemize}
    \item[A.] Fix $\sigma_B,\beta$ as constant parameters, and update $\rho_{RB}$ and dual variables $\Lambda_R$.
    \item[B.] Fix $\rho_{RB},\beta$ as constant parameters, and update $\sigma_B$.
    \item[C.] Fix $\rho_{RB},\sigma_B$ as constant parameters, and update $\beta$.
\end{itemize}
A notable advantage of the algorithm is that the update of primal variables $\rho_{RB}$ and dual variables $\Lambda_{R}$ is in closed form, guaranteed by Lemma \ref{SDP_lemma}; when $\rho_{RB},\sigma_B$ are fixed, the dual variable $\beta$ can be updated efficiently via a one-dimensional monotone equation with only a few iterations thanks to Lemma~\ref{lem:beta}.

\subsection{Updating $\rho_{RB}$ and Dual Variables $\Lambda_{R}$}
First, we minimize the Lagrangian function \eqref{Lagrangian} with respect to $\rho_{RB}$ as follows:
\begin{equation*}
        \min_{\rho_{RB}\in \mathbb{H}_{+}^{mn}}\tr(\rho_{RB}\log \rho_{RB})
    -\tr(\rho_{RB}\Delta').
\end{equation*}
Here, 
\begin{equation*}
    {\Delta}'=-\beta \Delta+\Lambda_R\otimes I_m+I_n\otimes\log \sigma_B,
\end{equation*}
and it is easy to verify that $\Delta'$ is an $(mn) \times (mn)$ Hermitian matrix. 
The following lemma provides a closed form solution to the above problem.
It solves {\color{black}an optimization problem} that is similar to that in the Gibbs' variational principle (cf. Lemma 1.1 in~\cite{Ramakrishnan2021Computing}), but without an additional constraint $\tr \rho = 1$ for the variable $\rho$. 
\begin{lemma}\label{SDP_lemma}
    Letting $\Tilde{\Delta}\in \mathbb{H}^n$, then the following \textcolor{black}{program with a semi-definite constraint}
    \begin{equation*}
        \min_{\rho\in \mathbb{H}^{n}_{+}}\tr(\rho\log \rho)
    -\tr(\rho\Tilde{\Delta}),
    \end{equation*}
    has the only solution $\rho=\exp(\Tilde{\Delta}-I_n)$, where we recall that $\exp(\cdot)$ is the matrix exponential function and $I_n$ is the $n \times n$ identity matrix.
\end{lemma}
\begin{IEEEproof}
    Denoting $\Tilde{\rho}=\rho/\tr( \rho)$, then the objective function can be written as 
    \begin{multline}\label{two_part}
        \tr (\rho) \left(\tr(\Tilde{\rho}\log \Tilde{\rho})-\tr\left(\Tilde{\rho}\log \frac{\exp(\Tilde{\Delta})}{\tr (\exp(\Tilde{\Delta}))}\right)\right)\\
        +\left(\tr (\rho) \log \tr(\rho)-\tr (\rho) \log \tr(\exp(\Tilde{\Delta})\right).
    \end{multline}
    For the first part in \eqref{two_part}, 
    \begin{equation*}
        \tr(\rho)\! \left(\!\tr(\Tilde{\rho}\log \Tilde{\rho})\!-\!\tr\left(\Tilde{\rho}\log \frac{\exp(\Tilde{\Delta})}{\tr (\exp(\Tilde{\Delta}))}\right)\!\right)\!=\!\tr (\rho) S(\Tilde{\rho}\|\Tilde{\sigma}),
    \end{equation*}
    in which $\Tilde{\sigma}=\frac{\exp(\Tilde{\Delta})}{\tr (\exp(\Tilde{\Delta}))}$. By Klein's inequality \cite{nielsen2010quantum}, 
    \begin{equation*}
       S(\Tilde{\rho}\|\Tilde{\sigma})\geq 0, 
    \end{equation*}
    where the equality holds if and only if $\Tilde{\rho}=\Tilde{\sigma}$, \textit{i.e.}, $$\rho/\tr( \rho)=\frac{\exp(\Tilde{\Delta})}{\tr (\exp(\Tilde{\Delta}))}.$$

    For the second part in \eqref{two_part}, denoting $t=\tr(\rho)$, then we have
    \begin{equation*}
        f(t)\triangleq t\log t-t\log \tr(\exp(\Tilde{\Delta})).
    \end{equation*}
    The convex function $f(t)$ achieves its minimum at $$t=\frac{1}{e} \tr(\exp(\Tilde{\Delta})),$$ since $f^{\prime}(t)=0$.

    Combining the two parts, we obtain 
    \begin{equation*}
        \begin{aligned}
            \rho/\tr( \rho)&=\frac{\exp(\Tilde{\Delta})}{\tr (\exp(\Tilde{\Delta}))}\\
            \tr(\rho)&=\frac{1}{e} \tr(\exp(\Tilde{\Delta})).
        \end{aligned}
    \end{equation*}
    Thus, the optimal $\rho$ is obtained as
    \begin{equation*}
        \rho=\exp(\Tilde{\Delta}-I_n).
    \end{equation*}
\end{IEEEproof}

Due to Lemma \ref{SDP_lemma}, $\rho_{RB}$ achieves the minimum at $\rho_{RB}=\exp(\Delta'-I_{mn})$, \textit{i.e.},
\begin{multline}\label{rho_RB}
    \rho_{RB}=\exp(I_n\otimes\log \sigma_B-\beta\Delta+\Lambda_R\otimes I_m-I_{mn}),
\end{multline}
in which $I_{mn}$ is the $mn \times mn$ identity matrix.

Substituting \eqref{rho_RB} into the constraint $\tr_2(\rho_{RB})=\rho_R$, we obtain the following equation of $\Lambda_R$, 
\begin{equation*}
    \rho_R=
    \tr_2(\exp(I_n\otimes\log \sigma_B
    -\beta\Delta+\Lambda_R\otimes I_m-I_{mn})).
\end{equation*}
Then we can update $\Lambda_R$ in the following way:
\begin{multline}
\label{Lambda_R}
    \exp(-\Lambda_R)\gets \exp(-\Lambda_R/2)\rho_R^{-1/2}K_0\rho_R^{-1/2}\exp(-\Lambda_R/2),
\end{multline}
where 
$$K_{0}= \tr_2\left(\exp(I_n\otimes\log \sigma_B-\beta\Delta+\Lambda_R\otimes I_m-I_{mn}\right))
$$
and $\rho_R^{-1/2}$ is the matrix square root of $\rho_R^{-1}$.

\begin{remark}
If $\rho_{R}$ is not invertible or it is ill-conditioned, we can add a  regularized term and update $\Lambda_R$ as follows:
\begin{equation}
\label{alpha}
\begin{aligned}
    \exp(-\Lambda_R)\gets \exp(-\Lambda_R/2) (\rho_R+ \alpha I_n)^{-1/2}
    \\
    (K_0+ \alpha I_n)(\rho_R+ \alpha I_n)^{-1/2}\exp(-\Lambda_R/2).
\end{aligned}
\end{equation}
Here $\alpha>0$ is a small regularity parameter. 
\end{remark}

\subsection{Updating Variables  $\sigma_B$}
Next, minimizing the Lagrangian function \eqref{Lagrangian} with respect to $\sigma_{B}$, we obtain the following problem:
\begin{equation}\label{sigmaB}
    \min_{\sigma_B\in \mathbb{D}^{m}} -\tr(\tr_1(\rho_{RB})\log \sigma_B).
\end{equation}
By Klein's inequality \cite{nielsen2010quantum},  we have $S(\tr_1(\rho_{RB})\|\sigma_B)\geq 0$.
Hence the lower bound for \eqref{sigmaB} is obtained as
\begin{equation*}
    -\tr(\tr_1(\rho_{RB})\log \sigma_B)\geq -\tr(\tr_1(\rho_{RB})\log \tr_1(\rho_{RB})),
\end{equation*}
and the equality holds if and only if $\sigma_B=\tr_1(\rho_{RB})$. Thus,
\begin{equation*}
       \sigma_B=\tr_1(\rho_{RB}),
\end{equation*}
is the minimizer.

\subsection{Updating Dual Variable $\beta$}
{Recall that it suffices to address the minimization in~\eqref{QRD} with the equality constraint $\tr(\Delta \rho_{RB})= D$.
Substituting \eqref{rho_RB} into the equality constraint,} we can obtain the following equation of $\beta$
\begin{equation}
\label{eq:originalroot}
    \tr\left(\exp(
    I_n\otimes\log \sigma_B
    -\beta\Delta+\Lambda_R\otimes I_m-I_{mn})\Delta\right)-D=0.
\end{equation}
Let
\begin{multline}
\label{Gbeta}
G(\beta)= \tr(\exp(
    I_n\otimes\log \sigma_B
    -\beta^{(k)}\Delta+\Lambda_R\otimes I_m-I_{mn})\\
    \exp(\beta^{(k)}\Delta)\exp(-\beta \Delta)\Delta)-D,
\end{multline}
where $\beta^{(k)}$ is the computed value of $\beta$ in the $k$-th iteration.
Then we can update $\beta$ in the following way:
\begin{equation}
\label{beta}
\beta^{(k+1)} \leftarrow \text{ the solution }\beta \in \mathbb{R}^+ \text{ of } G(\beta) = 0,
\end{equation}
i.e. $\beta^{(k+1)}$ is updated as the root of $G(\beta)$. 
The following lemma shows the function $G(\beta)$ in \eqref{Gbeta} is monotone in $\beta$, thus $\beta^{(k+1)}$ can be solved by Newton's method for a single variable function.

\begin{lemma}\label{lem:beta}
    The function $G(\beta)$ in \eqref{beta} is monotone in $\beta$.
\end{lemma}
\begin{IEEEproof}
Denote $$A=\exp(
    I_n\otimes\log \sigma_B
    -\beta^{(k)}\Delta+\Lambda_R\otimes I_m-I_{mn}),$$
    which is a positive semi-definite matrix. Then $G(\beta)$ can be written as
    \begin{equation*}
        G(\beta)=\tr(A
    \exp(\beta^{(k)}\Delta)\exp(-\beta \Delta)\Delta)-D.
    \end{equation*}
    The derivative of $G(\beta)$ can be calculated as
\begin{equation*}
\begin{aligned}
G'(\beta)=&-\tr(A\exp(\beta^{(k)}\Delta)\exp(-\beta \Delta)\Delta^2)\\
    =&-\tr(A\Delta\exp((\beta^{(k)}-\beta)\Delta)\Delta).
\end{aligned}
\end{equation*}
Here, the second equality is due to $\exp(\beta^{(k)}\Delta),\exp(-\beta \Delta),\Delta$ commutate with each other.
Since $\Delta\exp((\beta^{(k)}-\beta)\Delta)\Delta$ is also a positive semi-definite matrix and the trace of the multiplication of two positive semi-definite matrices is non-negative, we have $G'(\beta)\leq 0$. Hence, $G(\beta)$ is monotone in $\beta$.
\end{IEEEproof}

\subsection{Implementation}
The overall algorithm is summarized as Algorithm~\ref{alg1}. Here $\lambda_{\min}\left(\Delta_B\right)$ in line 3 can be efficiently computed by the inverse iteration method~\cite{Golub2013}.
A notable advantage of Algorithm~\ref{alg1} is that all variables except $\beta$ are updated in closed form.
Moreover, the dual variable $\beta$ can also be updated via Newton's method with only a few iterations, in light of Lemma~\ref{lem:beta}. 

\begin{figure}[!t]
\begin{algorithm}[H]
\caption{Alternating Minimization Algorithm}
\label{alg1}
\begin{algorithmic}[1]
    \REQUIRE Distortion threshold $D$, input quantum state $\rho_R$.
    \ENSURE Minimum $\tr(\rho_{RB}\log \rho_{RB})    -\tr(\tr_1(\rho_{RB})\log \sigma_B)-\tr(\rho_R\log \rho_R)$. 
    \STATE \textbf{Initialization:} $\sigma_B=\frac{1}{n}\mathbf{1}_{n},\ \Lambda_R=\mathbf{0}_{m},\ \beta=1$.    {
    \STATE $\Delta_B \leftarrow \tr_1(\Delta(\rho_R \otimes I_m))$.
    \IF{$\lambda_{\min}\left(\Delta_B\right) \leq D$}\RETURN minimum $0$.
    \ENDIF}
    \FOR{ $k = 1 : max\_iter$}
    \STATE Update $\Lambda_R$ as \eqref{Lambda_R}.
    \STATE $\rho_{RB}\leftarrow\exp(I_n\otimes\log \sigma_B-\beta\Delta+\Lambda_R\otimes I_m-I_{mn})$.
    \STATE Update $\sigma_B\leftarrow \tr_1(\rho_{RB})$.
    \STATE Update $\beta$ as \eqref{beta} by Newton's method.
    \ENDFOR
    \RETURN 
\end{algorithmic}
\end{algorithm}
\end{figure}




\begin{remark}
If the distortion matrix  $\Delta$ is the entanglement fidelity distortion, then the symmetry reduction acceleration technique in~\cite{he2024efficient} can be integrated into our algorithm as well.
Furthermore, the computational cost of each iteration in Algorithm~\ref{alg1} is reduced from $O(n^6)$ to $O(n^3)$. 
The key is that after diagonalizing the input state $\rho_R$, a solution pair $(\rho_{RB},\sigma_B; \beta,\Lambda_R)$ exists with both $\Lambda_R$ and $\sigma_B$ diagonal, and~$\rho_{RB}$ belonging to a sparse matrix  space $\mathcal{V}_{sym}^n$.
Based on this, lines 7-10 in Algorithm~\ref{alg1} are sped up as follows.
\begin{enumerate}[i)]
    \item Matrices in $\mathcal{V}_{sym}^n$ are stored as sparse matrices and multiplication and addition for them are computed under sparse matrix operations (lines 7-8). 
    \item The eigen decomposition for an $n^2 \times n^2$ matrix in $\mathcal{V}_{sym}^n$ can be computed with $O(n^3)$ operations. Correspondingly, the exponential and logarithmic functions for matrices in $\mathcal{V}_{sym}^n$ can be evaluated using $O(n^3)$ operations (lines 7-8).
    \item Partial traces of matrices in $\mathcal{V}_{sym}^n$ are known to be diagonal. Off-diagonal entries are no longer computed (lines 7,9). 
    \item  Update $\beta$ by $\beta^{(k+1)} = \beta^{(k)} + \log \left(\frac{G(\beta^{(k)})+D}{D}\right)$ (line 10). 
\end{enumerate}
\end{remark}



\begin{remark}
In~\cite{wu2022communication,chen2023constrained}, algorithms were proposed to solve the classical rate-distortion problem under a fixed distortion criterion. 
Algorithm~\ref{alg1} extends algorithms therein to the quantum setting, where the classical scenario emerges as a special case as all matrices reduce to diagonal ones and thus, commuting with each other.
However, the quantum setting introduces significant complexities due to non-commutativity, necessitating a more intricate approach.
We establish Lemma~\ref{SDP_lemma} to ensure the closed form iteration and devise a sophisticated update mechanism to handle the challenges unique to the quantum setting.
This highlights the innovation we have made in generalizing methods for the classical setting to the quantum setting.
\end{remark}

\begin{remark}
The convergence analysis of Algorithm 1 presents a significant challenge. 
Specifically, to guarantee convergence, the variables $\beta$ and $\Lambda_R$ must simultaneously satisfy conditions~\eqref{rho_RB} and~\eqref{eq:originalroot} throughout the iterative process. This requirement is essential for preserving the descending property of the objective function, which involves solving a high-dimensional nonlinear equation---a task that is inherently complex.
In Algorithm 1, $\beta$ is updated via equation by~\eqref{beta}, while $\Lambda_R$ is updated using equation~\eqref{Lambda_R} in an alternating manner. However, it is not theoretically guaranteed that both conditions~\eqref{rho_RB} and~\eqref{eq:originalroot} can be satisfied simultaneously. Despite this limitation, numerical simulations indicate that Algorithm 1 converges to a KKT point of the problem~\eqref{QRD}. Given the convexity of the problem~\eqref{QRD}, this convergence behavior suggests that the algorithm is capable of computing an optimal solution.
\end{remark}

\section{Numerical Experiments}
This section is aimed at showing the performance of our algorithm by numerical examples.
All the experiments are conducted on a PC with 16G RAM 
and one Intel(R) Core(TM) i7-7500U CPU @2.70GHz using MATLAB. 

We compute the quantum rate-distortion functions $R(D)$ with different input states $\rho_R$, and the distortion is measured by the entanglement fidelity.
Specifically, consider the following two examples.
\begin{enumerate}
\item The input state $\rho_{R} = \frac{1}{n}I_n$ is uniform. In this case, the quantum rate-distortion function has an analytical solution as follows (cf. Theorem~11 in~\cite{Wilde2013quantum} or Theorem~4.16 in~\cite{he2024efficient}).
\begin{multline}
\label{eq:analytical}
    R(D) = \mathds{1}_{0 \leq D < 1-\frac{1}{n^2}}\cdot 
    \bigg[ \log (n^2)
    \\
    -H\left(1-D,\frac{D}{n^2-1},\frac{D}{n^2-1},...,\frac{D}{n^2-1}\right) \bigg].
\end{multline}
\item The input state is $\rho_{R} = \frac{XX^{H}}{\tr(XX^{H})}$, where $X$ is an $n \times n$ Gaussian random matrix. That is, the real and imaginary parts of each entry of $X$ follow the standard normal distribution $\mathcal{N}(0,1)$. Then $\rho_R$ is sampled from the Hilbert-Schmidt ensemble, which is unitarily invariant and has been widely studied in quantum information theory~\cite{Zyczkowski2011generating,Braunstein1996geometry}. 
\end{enumerate}

\subsection{Algorithm Verification and Comparisons with Other Algorithms}
\label{subsec:comparison}

\begin{figure}[!t]
    \centering
    \includegraphics[scale = 0.32]{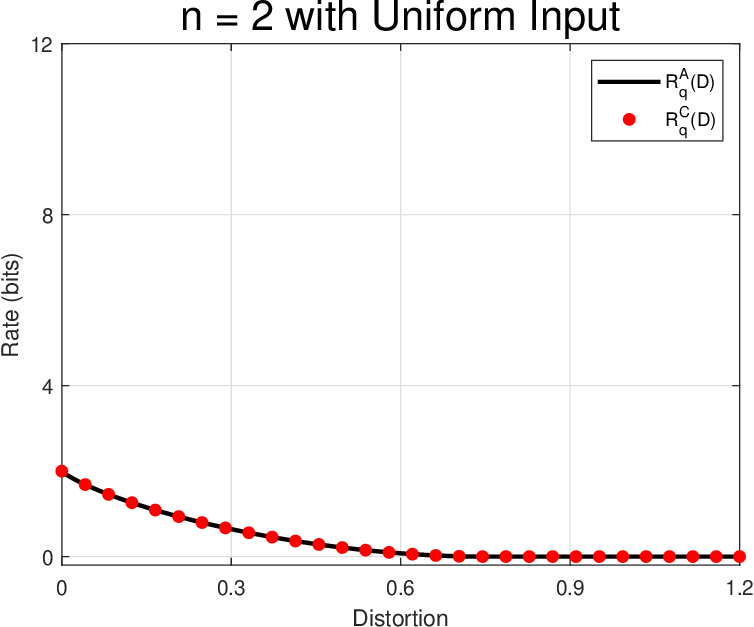}
    \includegraphics[scale = 0.32]{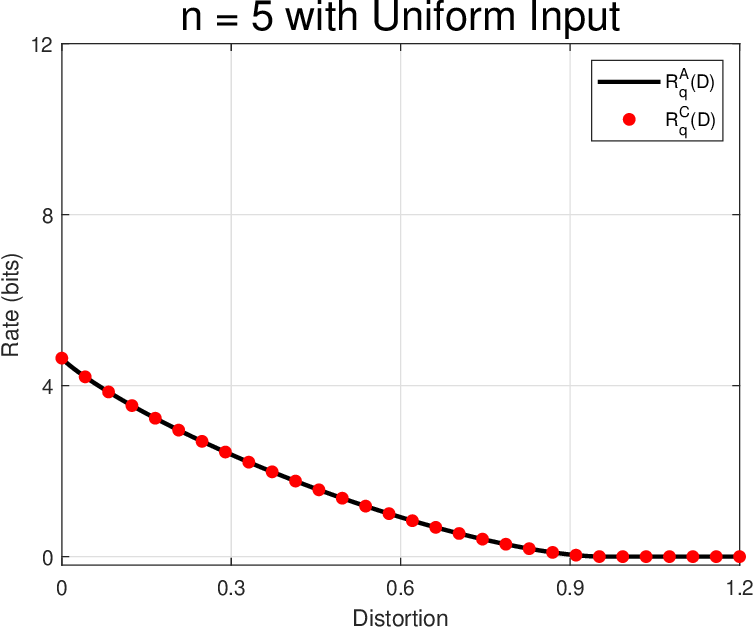}
    \includegraphics[scale = 0.32]{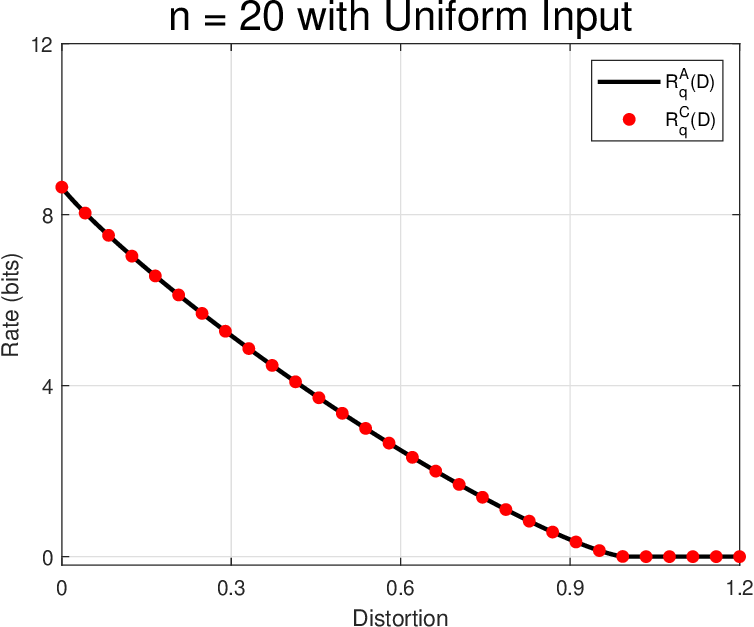}
    \includegraphics[scale = 0.32]{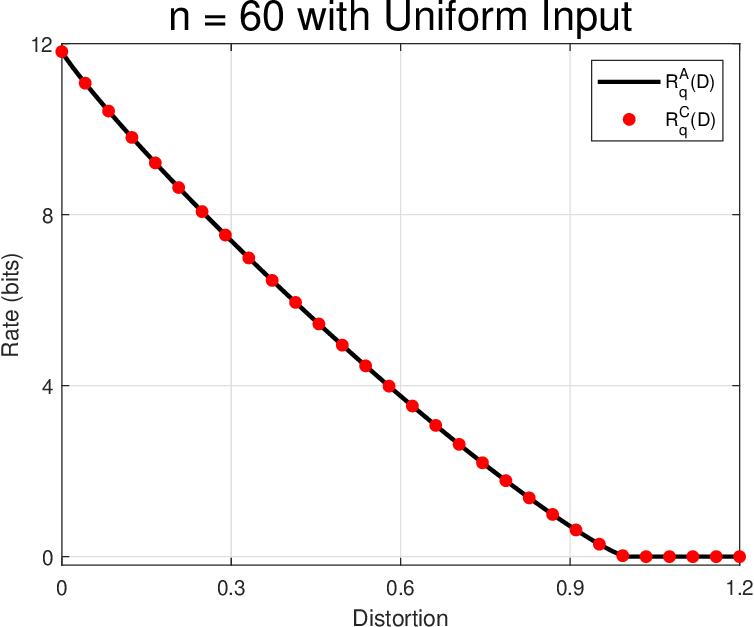}
    \caption{Analytical (black curve) and Numerical results (red dots) for the uniform input with $n = 2,5,20,60$. In each case, the algorithm computes $R(D)$ for $30$ consecutive points $D \in [0,1.2]$. }
    \label{fig:anyvsnum}
\end{figure}

For the first example, Algorithm 1 computes $R(D)$ for $n = 2,5,20,60$. The points generated by the algorithm are plotted in Fig.~\ref{fig:anyvsnum}, as well as the analytic curve given by the expression~\eqref{eq:analytical}.
Note that in Fig.~\ref{fig:anyvsnum}, the points generated by our algorithm lie exactly on the analytical curves in all considered cases, demonstrating the accuracy of our algorithm.

We compare our alternating minimization algorithm (AM) with the mirror descent algorithm (MD) in~\cite{he2024efficient}.  
{The EM algorithm in~\cite{hayashi2023bregman} requires solving a high-dimensional nonlinear system (of size $n^2$) for the update of $\rho_{RB}$, which presents significant computational challenges. 
More importantly, that work neither provides implementation details for solving the system nor presents numerical results demonstrating its practical performance.
Due to this lack of reproducible implementation and empirical validation, we omit comparisons with the EM algorithm~\cite{hayashi2023bregman}.}

We measure the performance of our AM algorithm and  the MD algorithm, by comparing their computational time in Table~\ref{tab:compare2}.
{The MD algorithm cannot compute the problem for a given distortion $D$.} To do the comparison, the dual variable $\beta$ is computed by an adaptive search. Both algorithms are stopped when the solutions are computed with absolute residual error less than~$10^{-8}$.

\begin{table}[!t] 
\renewcommand\arraystretch{1.4}
\centering
\caption{Comparison of Computational Time between the AM algorithm and the MD algorithm} 
\setlength{\tabcolsep}{2.4 mm}{
\begin{tabular}{c|c|c|c|c}
\toprule
\multirow{2}{*}{} & \multirow{2}{*}{($D$, $\beta$)} & \multicolumn{2}{c|}{Time (s)}& \multicolumn{1}{c}{Speed-up} \\
\cline{3-5}
&  & $t_{AM}$ & $t_{MD}$ & Ratio \\ 
\hline 
$n=20$ with & $(0.2,7.3753)$ & $0.0160$ & $1.4589$ & $91.0$ 
\\
Uniform Input	& $(0.7,5.1417)$ & $0.0534$ & $1.0533$ & $19.7$ 
\\
\hline 
$n=60$ with & $(0.2,9.5747)$ & $0.0861$ & $22.8500$ & $265$
\\
Uniform Input	& $(0.7,7.3411)$ & $0.3428$ & $17.2663$ & $50.4$ 
\\
\hline 
$n=20$ with & $(0.1,7.4391)$ & $0.0431$ & $2.9525$ & $68.5$
\\
Random Input	& $(0.3,3.8489)$ & $0.0750$ & $2.6542$ & $35.4$ 
\\
\hline 
$n=60$ with & $(0.1,9.3303)$ & $0.3536$ & $55.0462$ & $156$
\\
Random Input	& $(0.3,4.5368)$ & $0.6394$ & $44.4288$ & $69.8$ 
\\
\hline 
$n=180$ with & $(0.1,11.1761)$ & $4.5731$ & $2776.8$ & $607$
\\
Random Input	& $(0.3,5.6312)$ & $8.1797$ & $2102.3$ & $257$ 
\\
\bottomrule
\end{tabular}}
 
\vspace{+.03in}

\footnotesize{{Notes: a) Column 3-4 list the average computing time over $50$ trials, and column 5 is the speed-up ratio between our AM algorithm and the MD algorithm in~\cite{he2024efficient}. b) The MD algorithm cannot compute the rate directly with a given $D$, and hence we perform binary search on the dual variable $\beta$ to ensure accuracy. It generally takes about $\log (\frac{1}{\epsilon})$ trials to search for a suitable dual variable $\beta$ and compute $R(D)$ to within an absolute error $\epsilon$. c) Both algorithms are stopped until the absolute error is less than $\epsilon = 10^{-8}$.}}
\label{tab:compare2} 
\end{table}

From Table~\ref{tab:compare2}, we can see that our AM algorithm is much faster than the MD algorithm  if the problem is computed to the same level of accuracy. The advantage of our algorithm becomes more remarkable as the size of the problem gets larger.

\subsection{Convergence Verification}
\label{subsec:convergence}
To measure the convergence of the solution generated by our algorithm, it suffices to pay attention to the  residual error of the optimality condition, with respect to the solution pair $(\rho_{RB},\sigma_B;\Lambda_R,\beta)$ generated by the algorithm. The residual error of the optimality condition is defined by
\begin{equation}
\begin{aligned}
    e_{opt} = &|\tr(\Delta \rho_{RB})- D|+\frac{\left\|\tr_2(\rho_{RB})-\rho_R\right\|_1}{n^2} \\
    +&\frac{\left\|\tr_1(\rho_{RB})-\sigma_B\right\|_1}{m^2} + \frac{\left\| \rho_{RB}-\exp(\Tilde{\Delta}-I_{mn})\right\|_1}{m^2n^2},
\end{aligned}
\end{equation}
where $\|M\|_1 \triangleq \sum |M_{ij}|$ for a matrix $M = (M_{ij})_{i,j}$.

\begin{figure}[!t]
    \centering
    \includegraphics[scale = 0.32]{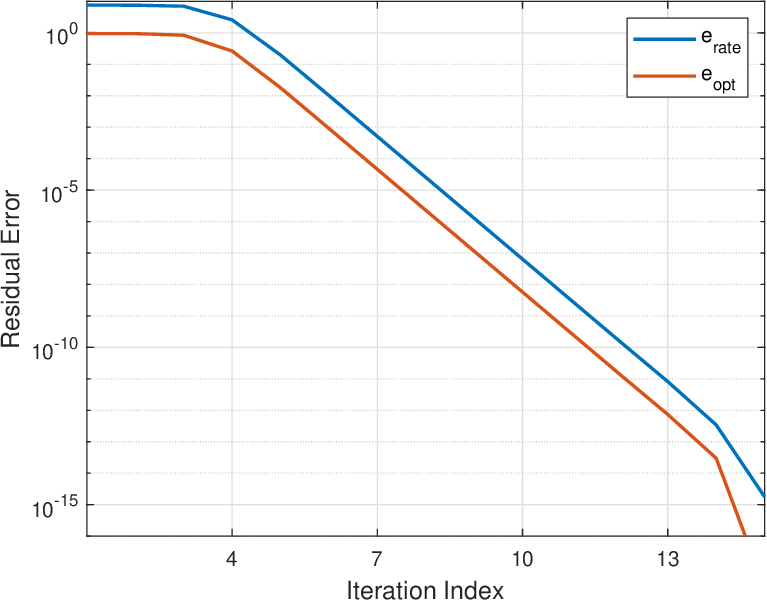}
    \includegraphics[scale = 0.32]{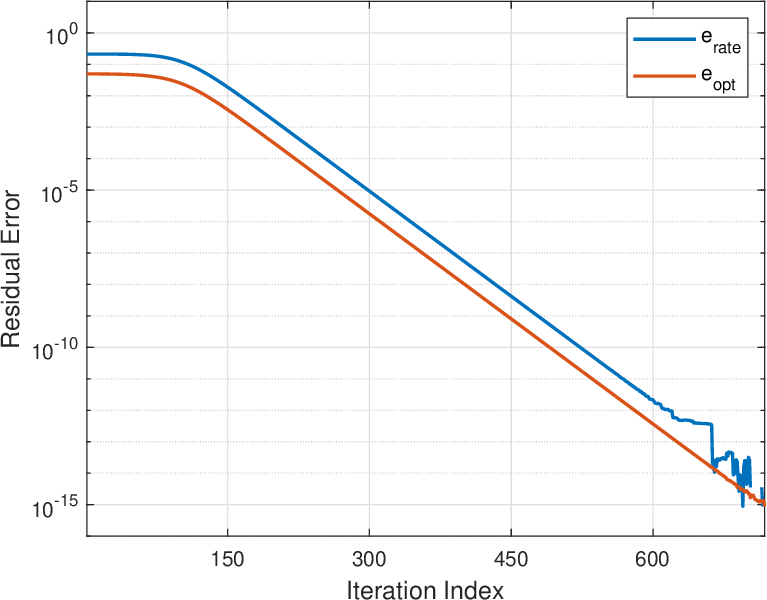}
    \includegraphics[scale = 0.32]{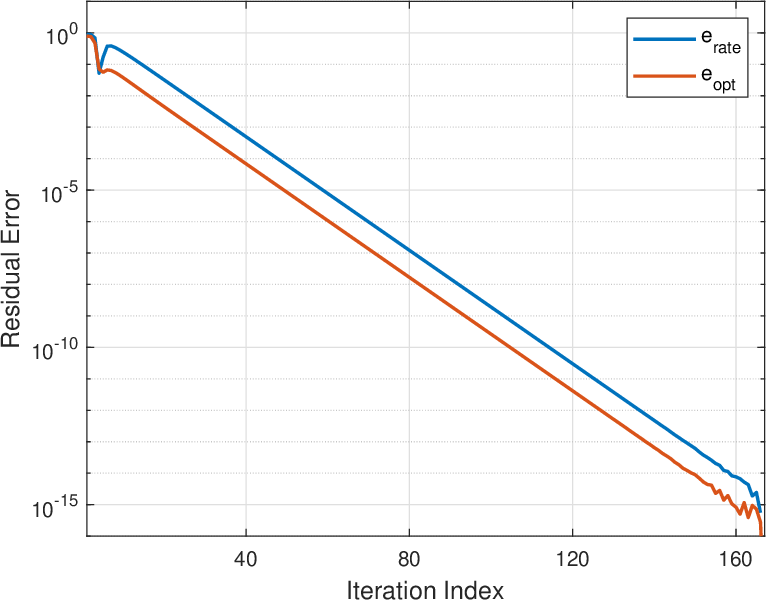}
    \includegraphics[scale = 0.32]{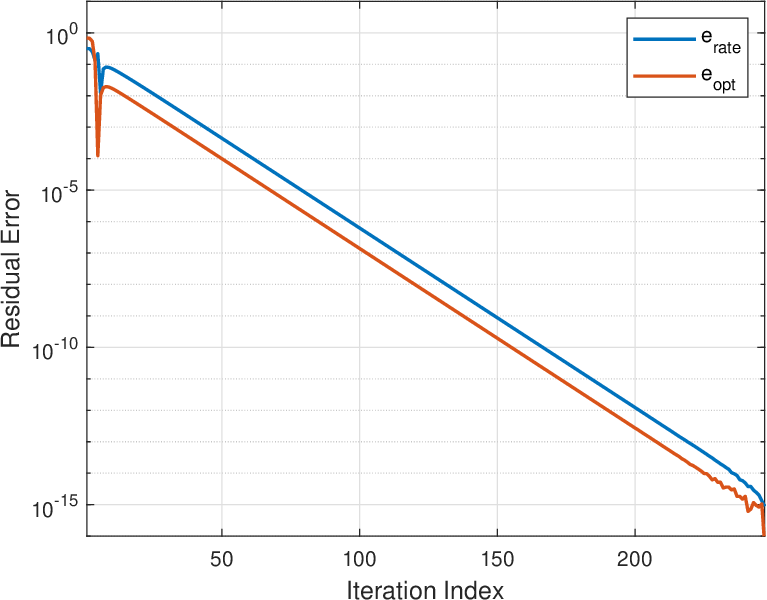}
    \caption{Convergence trajectories of the residual errors of the rates (blue) and the optimality conditions (orange) for $n = 60$ with the uniform input, where $D = 0.05$ (upper left), $D = 0.95$ (upper right), and the random input, where $D = 0.2$ (lower left), $D = 0.3$ (lower right).}
    \label{fig:optconver}
\end{figure}

The convergence trajectories of the residual errors of optimality conditions for both examples are plotted  in Fig~\ref{fig:optconver}. 
The algorithm is stopped until the residual error converges to the machine precision, i.e. $e_{opt} < 10^{-15}$. 
The convergence trajectories of the residual errors of the rates $e_{rate} \triangleq |S(\rho_{RB}\|\rho_R\otimes \sigma_B)-R(D)|$ are plotted in Fig~\ref{fig:optconver} as well. 
{\color{black}Here the reference value of $R(D)$ with the random input is computed as the output rate of the MD algorithm in~\cite{he2024efficient} after sufficiently many iterations.}
Both errors converge to the order $10^{-14}$ in all considered cases.
The convergence of our algorithm is hence verified.

{
\section{Conclusion}
In this work, we designed an efficient alternating minimization algorithm for computing the quantum rate-distortion function.
One advantage of our algorithm is that it solves the original problem rather than the Lagrangian relaxation of it, by updating the multiplier in each iteration.
The other is that all the other variables are iterated in closed form, thus avoiding solving multi-dimensional nonlinear
equations or multivariate optimization problems.
Through numerical experiments,  our algorithm was shown to achieve high accuracy and better computational efficiency relative to existing methods.
Developing effective computational methods for other theoretical bounds in quantum information theory is a possible direction for future works. 
}

\clearpage
\bibliographystyle{bibliography/IEEEtran}
\bibliography{bibliography/RD_REF}

\end{document}